%% file: aomartin_TC_Erice_notes.tex
\documentclass[11 pt, letterpaper]{revtex4}
\pdfoutput=1
\usepackage[dvips,letterpaper,text={6.5in,9in}]{geometry}
\usepackage{graphics}
\usepackage{amsmath}
\usepackage{colordvi}
\usepackage{epsfig}
\usepackage{latexsym}

\DeclareMathAlphabet\mathbb{U}{fplmbb}{m}{n}

\input{defs.tex}

\begin{document}
\title{Technicolor Signals at the LHC}
\author{Adam Martin}
\email{adam.martin@yale.edu}
\affiliation{Department of Physics, Yale University, New Haven CT, 06520}
\date{\today}
\begin{abstract}
In this lecture I give an introduction to technicolor and extended technicolor theories. I discuss the issues models of dynamical electroweak symmetry breaking struggle with and propose how non QCD-like dynamics, such as a `walking' or nearly-conformal technicolor coupling, can invalidate many conventional arguments against technicolor. I provide a short discussion of AdS/CFT-inspired extra-dimensional (Higgsless) models and their similarities and differences with older 4D technicolor models. To conclude, a summary of some possible LHC signatures is given.
\end{abstract}
\maketitle

\section*{Introduction}

Electroweak symmetry breaking is observed in nature: the weak bosons $W^{\pm}, Z^0$ are quite massive while the photon is massless. To remain consistent with gauge invariance, Lorentz invariance and unitarity, massive gauge bosons can only come about through spontaneously broken symmetry and the Higgs mechanism. Therefore, our electroweak theory is incomplete without some apparatus for spontaneous electroweak symmetry breaking (EWSB). One way to see this inconsistency is in the scattering amplitude of longitudinally polarized massive gauge bosons. These amplitudes violate unitarity at energy $\mathcal O(\tev)$ unless new contributions are added. Although we know spontaneous symmetry breaking must occur, the exact nature of its breaking is  unknown. I believe this is the most pressing question in High Energy Physics today and one that the LHC will hopefully answer. Until we have the answer, all possibilities should be explored and their similarities and differences fully fleshed out.

The setup of this lecture is the following: After briefly motivating extensions of the standard model and dynamical symmetry breaking (Section~\ref{sec:higgs}), I give an introduction to classic, rescaled-QCD technicolor (Section~\ref{sec:TC}) and its extension Extended Technicolor (Section~\ref{sec:ETC}). The hurdles which theories of dynamical electroweak symmetry breaking face are then laid out in Section 4. Many of these hurdles are based on the assumption that technicolor behaves just like QCD, and in Section 5 I illustrate how one example of non QCD-like behavior -- a nearly conformal or `walking' technicolor coupling constant -- can dramatically improve the situation. Other technicolor reviews which cover similar information are Ref.~\cite{Lane:1993wz, Lane:1994vu,Chivukula:1998if, Lane:2000pa, Lane:2002wv, Hill:2002ap}.  In Section 6 I investigate modeling dynamical electroweak symmetry breaking using an extra dimension.  Finally, collider signals for two different phenomenological models of technicolor are explored at length in Section 7.
\section{ Why do we need the Higgs?: Separating Fact and Fiction}
\label{sec:higgs}
 The simplest candidate for the Higgs mechanism is a single Higgs doublet whose potential is arranged by hand to have a minimum at non-zero vacuum expectation value. Although adequate, the simple Higgs scalar is theoretically unsatisfying. The mass parameters in the potential are sensitive to the highest momentum in the theory, $\Lambda_{UV}$. In order to preserve a hierarchy between $\Lambda_{UV}$ and the weak scale $\Lambda_{EW}$, an incredible  and `unnatural' degree of tuning is necessary. A second problem with the simple Higgs theory is triviality. Triviality is the statement that the cutoff of the theory cannot be taken arbitrarily high without the theory becoming free~\cite{Chivukula:1998if, Wilson:1971dh,Wilson:1973jj}. Therefore the simple Higgs doublet model must have a finite cutoff.  The straightforward way to see this is to look at the renomalization group equation for the Higgs quartic coupling.

To rectify these unsatisfactory aspects of the Higgs theory, we believe new physics is necessary.  The most advocated extension is supersymmetry, in which the naturalness problems is improved by adding a symmetry between bosons and fermions. As a result, each standard model (SM) particle has a superpartner with opposite statistics yet equal couplings. Supersymmetry accomplishes EWSB while  remaining weakly coupled, making a plethora of calculations -- spectrum, cross sections, decay rates, etc. all doable. This is certainly a nice feature, but it is by no means a requirement. Additionally, even the minimal viable model, the MSSM, leads to an improved unification of the gauge couplings and a dark matter candidate. While these `free' MSSM aspects are theoretically interesting, they are not unique, nor are they necessary for EWSB. Moreover, fundamental spin-0 fields -- which are a key ingredient in supersymmetry --  have never been observed in nature.  

\subsection{Dynamical Symmetry Breaking: Higgs mechanism without the Higgs}

The phase transitions analogous to EWSB which have been observed happen dynamically. Two obvious precedents are chiral symmetry breaking in QCD and superconductivity. 

In QCD, in the limit of massless $u,d$ quarks, the theory has an $SU(2)_L \otimes SU(2)_R \otimes U(1)_B$ global symmetry, where $L, R$ refer to rotations among the left-handed, right-handed chirality quarks. At low energy, to a good approximation, the states fall into representations of $SU(2)_V$ -- where both left-handed and right-handed fermions feel the same rotation, rather than the full symmetry. What has happened to the axial combination of the rotations? This symmetry is not manifest in the spectrum, thus it must be broken. However, since the theory only consists of massless quarks and gluons, we are forced to conclude that the symmetry is broken as a result of the strong interaction itself. A simple way to present this, consistent with the observed symmetry-breaking pattern, is to assume the bilinear fermion operator $q_L \bar q_R$ has obtained a  nonzero expectation value. This is often referred to as the fermions {\em forming a condensate}.

Dynamical symmetry breaking is also at work in superconductivity. Though the situation is quite different -- the system of interest is non-relativistic and the interactions are weak -- the properties of superconductors were explained by introducing a charge $-2$ composite of electrons~\cite{Bardeen:1957mv}. Therefore, as in QCD, the symmetry breaking is explained not by fundamental scalar fields, but by new `high-energy' dynamics.

In addition to having precedents in nature, dynamical symmetry breaking via asymptotically free gauge dynamics also explains large scale hierarchies. Given a UV scale and a coupling at that high scale, an exponentially smaller scale is automatically generated -- without fine tuning any parameters! This is the mechanism of dimensional transmutation.
\be
\label{eq:dimtrans}
\Lambda_{IR} \sim \Lambda_{UV}\exp{\Big( \frac{-8\pi^2}{b_0 g^2(\Lambda_{UV})}\Big)}  
\ee
The basic goal of technicolor~\cite{Weinberg:1979bn, Susskind:1978ms} theories is to break electroweak symmetry through some new asymptotically-free gauge interaction. We explore the basic setup in the next Section. Once dynamical electroweak symmetry breaking has been arranged, dimensional transmutation naturally explains why $\Lambda_{EW}$ is exponentially smaller than $\Lambda_{UV}$. No other extension of the standard model can explain the gauge hierarchy as simply or as naturally! However, just as in QCD, dynamical electroweak symmetry breaking necessarily leads us to strong coupling and incalculability.

\section{Classic Technicolor}
\label{sec:TC}
  Classic technicolor is heavily inspired by QCD. It was invented at a time when the heaviest fermion was the bottom quark, whose mass of $\sim 5\ \gev $ is much less than the weak scale.
%

The building blocks of technicolor are a set of massless fermions (technifermions) which feel a new non-abelian gauge interaction (technicolor) with new force-carrying fields (technigluons). We assume for simplicity that the technifermions are charged in the fundamental representation, though other representations can also be used. The left-handed components of the technifermions form electroweak doublets, while the right-handed components are electroweak singlets. Both left-handed and right-handed technifermions carry hypercharge. These charge assignments are summarized below:
\begin{gather}
T = \left( \begin{array}{c} U_L \\ D_L \end{array} \right) \in (N_{TC}, 2)_{Y_L}\nn \\
~U_R \in (N_{TC}, 1)_{Y_U}\ ,\quad  D_R \in (N_{TC}, 1)_{Y_D} \nn
\end{gather}
under $(SU(N_{TC}), SU(2)_w)_{U(1)_Y}$, where $SU(N_{TC})$ is the technicolor strong interaction. Ignoring electroweak interactions, the technicolor sector has a global symmetry $SU(N_{F})_L \otimes SU(N_{F})_R$ where $N_F$ is the number of techniflavors.
%

 We assume the technicolor interaction becomes strong at $\sim \Lambda_{EW}$, causing the formation of a techni-condensate:
\be
\langle U \bar U \rangle = \langle D \bar D \rangle = 4\pi F^3_T, {\rm where}\ \langle U \bar U \rangle  = 2 \langle U_L U^{\dag}_R \rangle.
\ee
As the left-handed technifermions carry electroweak quantum numbers while the right-handed technifermions do not, the formation of the techni-condensate breaks electroweak symmetry.  

If we temporarily shut off the gauge couplings, the pattern of global symmetry-breaking in the technicolor sector is $SU(N_F)_L \otimes SU(N_F)_R$ down to the diagonal combination $SU(N_F)_D$~\footnote{This often appears in the literature as $SU(2N_D)_L \otimes SU(2N_D)_R \rightarrow SU(2N_D)_D$ where $N_D$ labels the number of electroweak doublets}. The fact that the low-energy theory contains a residual (or `custodial') $SU(N_F)_D$ symmetry in the limit of zero gauge couplings is extremely important. Turning on the gauge interactions, this `custodial' symmetry forces the $W^{\pm}$ and $Z^0$ masses have the correct pattern (up to loop-level electroweak corrections):
\be
\label{eq:TCEWSB}
m^2_W \sim \frac{g^2\frac{N_F}{2}F^2_T}{4} = m^2_Z \cos^2{\theta},
\ee
where $N_F/2$ is the number of electroweak doublet technifermions.
For a technicolor theory with more than one electroweak doublet (2 techniflavors), the symmetry-breaking produces more Goldstone bosons than can be eaten by the gauge bosons. These extra states are pseudoscalars and are massless at tree level. These states are generally called {\em techni-pions} and we will return to the interactions which give them mass shortly.

Without the ability to calculate at strong coupling, the simplest way to estimate the properties of an EW-scale strong interaction is to scale what we know from QCD. These estimates are often accompanied by additional scalings to account for a different number of technicolors or techniflavors as compared to QCD. However, all these scalings are naive in that they have no first-principles motivation. We have already drawn the connection between the pions in QCD and the techni-pions -- including the longitudinal $W^{\pm},Z$ of technicolor. We can extend the analogy further, postulating spin-1 vector and axial isovector mesons $\rho_T, a_T$, and isosinglets $\omega_T$. The corresponding QCD states are the $\rho, a_1$, and $\omega$. In QCD the spin-1 states are narrow, making them easier to find experimentally, however we expect technicolor would also lead to spin-0 and spin-2 states as well. For a naive estimate the masses and widths, we simply rescale the QCD values by factors of $\frac{F_T}{f_{\pi}}, \frac{N_{TC}}{3}$ and $\frac{N_F}{2}$~\cite{Lane:2002wv}.
\bea
M_{\rho_T} \sim \sqrt{\frac{3}{N_{TC}}}\times 2\ \tev ,\quad \Gamma(\rho_T \rightarrow W_L W_L) \sim 500 \Big( \frac{3}{N_{TC}}\Big)^{3/2}\ \gev.
\eea

Given the many new states -- techni-vectors, techni-pions, etc. in a technicolor model, how will they be produced and observed at colliders like the LHC? So far we have not described any interaction between the technicolor states and the standard model quarks and leptons, thus new technicolor states must be produced through their interactions with gauge bosons. The simplest mechanism, known as vector meson dominance, is shown on the left below:
\begin{figure}[th!]
\centering
\begin{minipage}[c]{0.45\linewidth}
   \centering
   \includegraphics[width = 2.25 in, height=1.05 in]{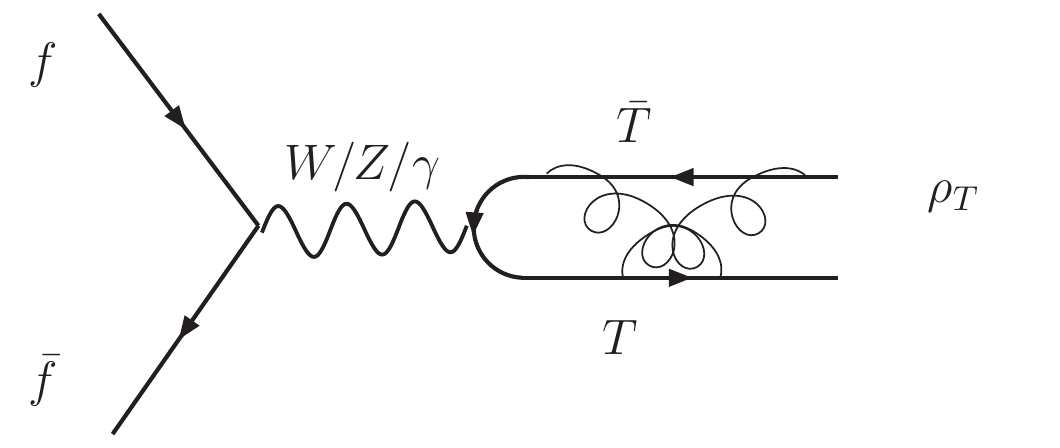} 
 \end{minipage}
\hspace{0.5 cm}
\begin{minipage}[c]{0.45\linewidth}
   \centering
   \includegraphics[width = 2.25 in, height=1.25 in]{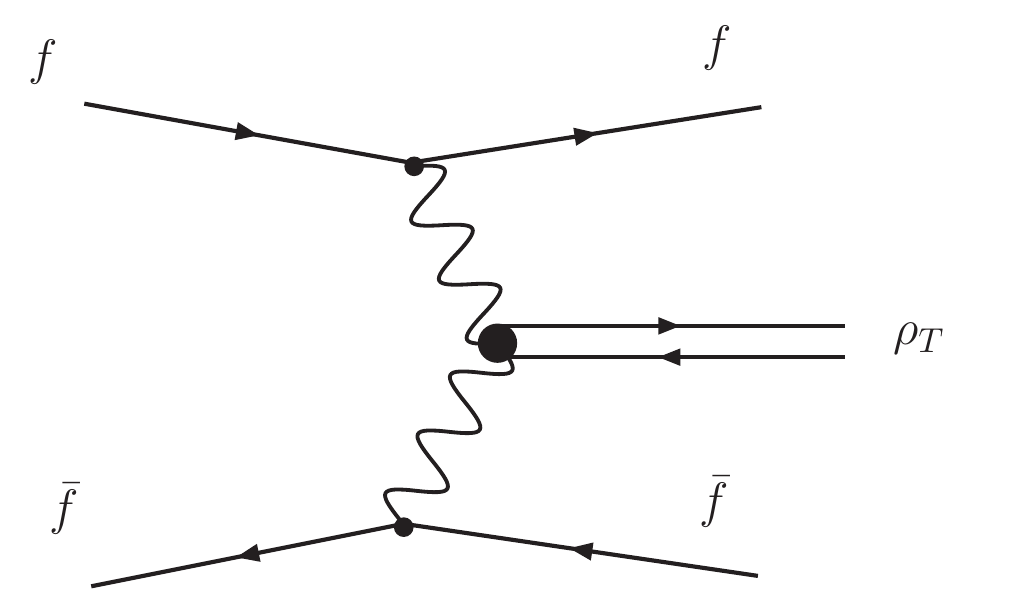}
\end{minipage}  
\end{figure}
A SM gauge boson, $\gamma, W^{\pm}, Z$ is produced from the collision of two fermions. This gauge boson feels the effects of technifermions, thus can mix into a bound state of the technifermions, leading to an effective $f \bar f \rightarrow \rho_T$ interaction. This is identical to the mechanism used to  describe QCD vector meson production in $e^+ e^-$ collisions. Another production mechanism, shown in the right-hand side of the above figure, is to produce the techniresonances by the fusion of gauge bosons which have been emitted by the colliding quarks. This mechanism is known as vector-boson fusion. Some early technicolor phenomenology was explored in Refs.~\cite{Eichten:1984eu, Bagger:1993zf, Bagger:1995mk}.

\section{Extending Technicolor}
\label{sec:ETC}
While the chiral symmetry breaking caused by technicolor provides the mechanism to give the $W^{\pm}, Z^0$ bosons mass, the technicolor interaction alone does not provide way to give mass to the standard model fermions. This crucial job is easily accomplished by the SM Higgs. However, to maintain a natural explanation of the weak scale (Eq.(\ref{eq:dimtrans})) we cannot use scalars and we must get fermion masses using gauge dynamics alone.

The most-studied method for accomplishing this feat is known as Extended Technicolor~\cite{Eichten:1979ah}. The basic postulate of ETC is that there is a new gauge interaction  -- named extended technicolor -- under which both fermions and technifermions transform. Putting SM fermions and technifermions into the same representations, a fermion can turn into a technifermion by emitting an ETC gauge boson. Since we see no other light gauge bosons (other than the SM gauge bosons), the ETC symmetry must be broken at some very high scale, giving the corresponding ETC gauge boson masses of the same order of magnitude. Ideally the breaking of ETC is also done dynamically, otherwise we have merely replaced the Higgs hierarchy problem for the ETC-hierarchy problem. Given the ambiguity about the structure and the severe constraints and functions ETC must fulfill, an exact model of ETC has so far been beyond our grasp. We must be content to play with the effective theory which results from integrating out the massive ETC bosons, leaving the full theory for the future. This is a bit of a cop-out, but no more than is necessary in SUSY or any other extension of the standard model.

Integrating out the massive ETC gauge bosons, we are left with a set of dimension six operators at the ETC energy scale~\cite{Hill:2002ap}: 
\be
\label{eq:abcETC}
g^2_{ETC}\Big(\alpha_{ab}\frac{(\bar T \gamma_{\mu}t^aT)(\bar T \gamma^{\mu}t^b T)}{M^2_{ETC}}\Big|_{ETC}
 + \beta_{\alpha\beta} \frac{(\bar T \gamma_{\mu}t^aT)(\bar q \gamma^{\mu}t^b q)}{M^2_{ETC}} \Big|_{ETC}
 + \gamma_{\alpha\beta} \frac{(\bar q \gamma_{\mu}t^a q)(\bar q \gamma^{\mu}t^b q)}{M^2_{ETC}} \Big|_{ETC}\Big)\;,
\ee
where $t^a, t^b$ label the ETC generators, $g_{ETC}$ is the ETC gauge coupling, and the `$|_{ETC}$' indicate that these operators are generated at the matching scale $M_{ETC}$. The field content of the operators indicates what role they play. The $\alpha_{ab}$ term contains only techni-fields and is responsible for giving mass to the techni-pions. The $\beta_{ab}$ term contains both technicolor and SM fermions, thus once Fierz rearranged leads to:
\be
\beta_{ab}\frac{g^2_{ETC}(\bar T_b T_b)(\bar q_a q_a)}{M^2_{ETC}} \longrightarrow \Big(\frac{\beta_{ab}g^2_{ETC}\langle \bar T T \rangle|_{ETC}}{M^2_{ETC}}\Big)\bar q q\;,
\ee
and hence to quark and lepton masses. The remaining term with coefficient $\gamma_{ab}$ contains only SM fields and will lead to potentially dangerous FCNC interactions. 

While the techni-bilinear $\langle \bar T T \rangle$ is fixed at the electroweak scale by $G_F$ the fermion mass estimates above depend on its value at the ETC scale. The two techni-bilinear values are related by the renormalization group equations (RGE). Ignoring all interactions except for the strong $SU(N_{TC})$, the RGE for $\langle \bar T T \rangle$ is~\footnote{Conventionally, following naive dimension analysis, $\Lambda_{TC} \cong 4\pi F_T$}:
\be
\langle \bar T T \rangle = \langle \bar T T \rangle \exp{\Big(  \int_{\Lambda_{TC}}^{M_{ETC}} \frac{d\mu}{\mu} \gamma_{m}(\mu) \Big)}\;,
\ee
where $\gamma_{m}$ is the anomalous dimension of $\langle \bar T T \rangle$. As technicolor is a strongly coupled theory, we cannot calculate this exactly, however we can get some idea by using QCD as an analogy. In  QCD and at lowest order in perturbation theory, $\gamma_m \sim \mathcal O(\alpha) \ll 1$. If these two rather drastic approximations -- QCD-like behavior and the adequacy of lowest order perturbation theory -- can be extended to technicolor, then the anomalous dimension is small and can be dropped. In that extreme case,
\be
\langle \bar T T \rangle\Big|_{ETC} \approx \langle \bar T T \rangle \Big|_{TC} \equiv 4\pi F^3_T\;.
\ee
Working with this QCD-inspired, small anomalous dimension assumption, we can plug in the techni-condensate into the $\beta_{ab}$ term to get an estimate of the SM-fermion masses it leads to:
\be
m_q, m_{\ell} \sim \frac{g^2_{ETC}}{M^2_{ETC}}\langle \bar T T \rangle|_{ETC} \sim  \frac{g^2_{ETC}}{M^2_{ETC}} (4\pi F^3_T).
\ee

\section{Obstacles to Overcome}
\subsection*{Flavor-Changing Neutral Currents (FCNC)}

Even at its inception it was recognized that ETC would be tightly constrained by flavor-changing neutral currents (FCNC)~\cite{Eichten:1979ah}. These FCNC come from the $\gamma_{ab}$ terms of (\ref{eq:abcETC}), and are most stringently constrained by $\Delta S = 2$ interactions in the Kaon system. The ETC-mediated effective interaction which contributes to $\Delta S = 2$ is~\cite{Lane:2002wv}:
\be
\frac{g^2_{ETC}\theta^2_{sd}}{M^2_{ETC}}(\bar s \Gamma_{\mu} d)(\bar s \Gamma^{\mu} d) + h.c.\nn,
\ee
where $\Gamma_{\mu}$ is some Dirac matrix structure, and the quark flavors $s, d$ have been selected by the ETC generators. We have included a mixing angle $\theta_{ds}$ which is presumably $\mathcal O(1)$. The $K-\bar K$ mass difference is constrained to be $\Delta m_K < 3.5\times 10^{-12}\ \mev$, which we can turn into a constraint on the ETC interactions~\cite{Lane:2002wv}:
\be
\label{eq:FCNC1}
\frac{M_{ETC}}{g_{ETC}\sqrt{Re(\theta^2_{ds})}} \ge 1.3 \times 10^3\ \tev.
\ee
The constraints on the  imaginary part of the $K -\bar K$ mass difference, $\epsilon_K$ are tighter by three orders of magnitude, implying~\cite{Lane:2002wv}
\be
\label{eq:FCNC2}
\frac{M_{ETC}}{g_{ETC}\sqrt{Im(\theta^2_{ds})}} \ge 1.6 \times 10^4\ \tev.
\ee
Using ETC-scale parameters which pass the FCNC constraints, the resulting quark and lepton masses are forced to be very small,
\be
m_q, m_l \sim \frac{0.1\ \mev}{\Big(\frac{N_F}{2}\Big)^{3/2} |\theta_{ds}|^2}.
\ee
Unless we are willing to fine-tune the mixing angles, $\mathcal O(\gev)$ quark are masses difficult to achieve, and the top mass looks impossible.

\subsection*{Technicolor and Precision Electroweak}

While FCNC are a problem of ETC and not of fundamental technicolor, that does not make technicolor problem-free. Technifermions, by definition and necessity, communicate with the EW gauge bosons. The properties of the electroweak sector have been measured to extreme accuracy at experiments such as LEP and SLD and thus the a priori strong contributions to electroweak observables are tightly constrained~\cite{Peskin:1990zt, Golden:1990ig}.
\begin{figure}[ht!]
\centering
\includegraphics[height=0.75in, width=2.5in]{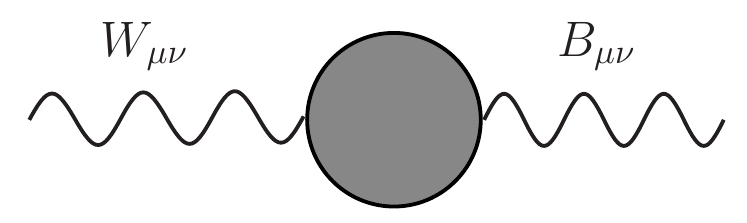}
\caption{A typical technicolor contribution to precision electroweak quantities. The blob represents an incalculable quagmire of technifermions and technigluons.}
\label{fig:TCblob}
\end{figure}
Assuming the contributions of new physics are felt only in the EW gauge sector -- an assumption which is certainly valid up to ETC scale effects in classic technicolor -- all deviations in the EW sector from tree level SM can be parameterized in terms of three quantities called S, T, and U. Precision measurements, taken together, restrict the allowed region of S,T, and U.

Deviations from SM tree level values come from two sources: loop level SM effects and tree-level new physics effects. The former contain divergences which are cut off by the Higgs mass, while the latter do not. However, to constrain new physics, one typically subtracts the SM-loop contribution from the extracted value of S,T,U. To maintain a finite observable the new physics contribution must inherit a piece which depends on the Higgs mass -- whatever is necessary to cancel the SM loop part. Unfortunately,  this often-confusing procedure results in an odd, residual scale dependence in the new physics contribution.

As we are unable to calculate the blobs in figure~(\ref{fig:TCblob}), we must make some assumption. The assumption made in the classic paper by Peskin and Takeuchi~\cite{Peskin:1990zt} is that the vector and axial current correlation functions which define $S$ are each saturated by a single, narrow resonance. Rescaling with colors and flavors, the result is:
\begin{gather}
\label{eq:Slab}
S = -4\pi\frac{d}{dq^2}\Big(\Pi_{VV}(q^2) -\Pi_{AA}(q^2) \Big)|_{q^2 = 0} \nn \\
{\rm single\ resonance:\ } S_{TC} = 4\pi\Big( 1 + \frac{M^2_{\rho_T}}{M_{a_T}}\Big) \frac{F^2_T}{M^2_{\rho_T}}\cong 0.25\frac{N_C}{3} \frac{N_F}{2},
\end{gather}
where $M_{\rho_T} (M_{a_T})$ is the mass of the first vector (axial) techni-resonance. This value is difficult to reconcile with the current PEW best fit: $S = -0.10 \pm 0.10$~\cite{Amsler:2008zz} (setting reference Higgs mass to $117\ \gev$ -- the limits are more negative if we increase the reference scale). QCD-like technicolor, especially with large values for $N_C$ (large $N_F$ too, though this is a subtler issue as we will see in the next section) appear to be ruled out. However, there is no first-principles reason why the current correlation functions in all strongly coupled theories are dominated by a single resonance. Such saturation does happen in QCD, but there is no reason to believe it is generic.

\section{Non QCD-like Technicolor Dynamics: Walking Technicolor}

A common assumption in the last section was that the technicolor dynamics could be adequately represented by rescaling QCD. While QCD is a useful  (and our only full) probe of strong dynamics, there is no reason to believe it is typical. However, playing devil's advocate, one may ask:  what argument is there that a strongly coupled theory can be very different from QCD? Consider changing the number of flavors in QCD. At 2 or 3 flavors we have QCD as we know it, however by looking at the first terms in the beta function we get the sense that this QCD-like behavior cannot persist for all $N_F$. Setting $N_C = 3$:
\begin{gather}
\beta(\alpha_T) = -2\beta_0 \frac{\alpha^2_T}{4\pi} - 2\beta_1 \frac{\alpha^3_T}{(4\pi)^2} + \cdots \\
{\rm for\ fundamentals: \ }\beta_0 = 11 - \frac{4}{3} \frac{N_F}{2},\ \qquad \beta_1 = 102 - \frac{38}{3}N_F.
\end{gather}
Eventually, at $N_F \approx 16$, the $\beta_0$ coefficient becomes negative indicating a loss of asymptotic freedom. A non-asymptotically free theory is not what we want for technicolor, however at intermediate values of $N_F$ the theory may have drastically different properties than in 2 or 3 flavor QCD. For example, at $N_F \sim 12$, one can balance the $\beta_0$ and $\beta_1$ terms, leading to a zero in the beta function to order $\alpha_T^4$. The value of $\alpha_T$ where this vanishing occurs, $\alpha^*_T$ is known as an interacting fixed point, since a vanishing beta function means $\alpha^*_T$ stays nearly fixed over wide range of scales~\footnote{For 12 flavor QCD, the value one gets for the interacting fixed point coupling using perturbation theory is $\alpha^*_T = \frac{6\pi}{25}$.}. This behavior is clearly distinct from a QCD-like beta function. While the complete details of a theory with slowly running, yet strong coupling are, so far, incalculable, rescaling QCD is not necessarily a good approximation. Attempts to study the behavior of gauge theories with different numbers of fermion flavors using lattice techniques, are currently underway~\cite{Appelquist:2007hu,Catterall:2007yx, Deuzeman:2008da,Fodor:2008hn, Fodor:2008hm}. The idea of a slowly evolving technicolor coupling is known as Walking Technicolor~\cite{Holdom:1981rm, Appelquist:1986an, Yamawaki:1986zg}. In the next section we work out some of the ramifications of a walking coupling on the low-energy theory.

\subsection*{Walking and FCNC:} 

The entire line of logic that lead to the FCNC problem started with the assumption that the anomalous dimension of the technifermion condensate was small, $\mathcal O(\alpha_{TC})$ in analogy with QCD. Relaxing this assumption, we can ask what happens if the anomalous dimension is large, order $1$. In that case, the condensate at the ETC scale, the scale relevant for determine fermion masses, is not the same as the TC scale condensate.
\begin{gather}
\label{eq:walkmq}
\langle \bar T T \rangle_{ETC} = \langle \bar T T \rangle_{TC}\exp{\Big( \int_{\Lambda_{TC}}^{M_{ETC}} \frac{d\mu}{\mu}(1 + \mathcal O(\alpha_T) \Big)  } \sim \langle \bar T T \rangle_{TC}\frac{M_{ETC}}{\Lambda_{TC}},\\ 
{\rm thus}\quad m_{q},m_{\ell}  \simle \frac{g^2_{ETC}}{M^2_{ETC}}(4\pi F^3_T)\Big( \frac{M_{ETC}}{\Lambda_{TC}}\Big) \sim \mathcal O({\rm few}\ \gev)\;.
\end{gather}
While this result isn't perfect, second generation fermion masses and possibly the $b$ mass now look tractable. The tecnipion masses, caused by the $\alpha_{ab}$ term of Eq.(\ref{eq:abcETC}) are also enhanced by walking,
\be
\label{eq:walkmpi}
m_{\pi_T} \sim \frac{g_{ETC}\langle \bar T T \rangle_{TC}}{F_T \Lambda_{TC}}.
\ee
One consequence of enhanced techni-pion masses which we will visit later is that the decays of technivector mesons to techni-pions, $\rho_T \rightarrow \pi_T \pi_T$, etc. may be kinematically closed.

 While the enhancement one gets from a large anomalous dimension definitely eases the tension between FCNC and reasonable quark masses, how does one actually calculate the anomalous dimension in a strongly coupled theory? One technique employed early on to calculate the anomalous dimension in walking TC theories is the rainbow approximation of the Schwinger-Dyson equation (SDE)  for the technifermion propagator~\cite{Pagels:1974se, Fukuda:1976zb, Higashijima:1983gx}. The SDE are equations of motion for operators and extend beyond perturbation theory. However, to get a simple expression, typically one truncates the perturbation expansion at lowest order, while simultaneously neglecting the effects of the technicolor interaction on the technigluon propagator. With these assumptions, one indeed finds the anomalous dimension for the technifermion bilinear to be large.
 \be
 \label{eq:SDE}
 \gamma_{m, SDE} = 1 - \sqrt{1 - \frac{3C_2(r)\alpha(\mu)}{\pi}} \rightarrow 1\ {\rm as\ } \alpha(\mu) \rightarrow \frac{\pi}{3C_2(r)} \equiv \alpha_c,
 \ee
  where $\alpha_c$ is interpreted as the critical coupling necessary for chiral symmetry breakdown.
  This expression does reduce to the conventional form at small coupling, however the assumptions involved are severe. Attempts to improve the SDE analysis, both analytically and numerically have been attempted, with some evidence that the lowest order analysis is stable~\cite{Appelquist:1988yc, Cohen:1988sq, Kurachi:2006ej}. While the exact value of the anomalous dimension in Eq. (\ref{eq:SDE}) may be debatable, it is clear that large anomalous dimensions are possible in strongly coupled theories and can significantly improve the FCNC situation.
  
\subsection*{Walking and the $S$ parameter}
\label{sec:PEWwalk}

The effects of walking on the precision electroweak parameters are more speculative. At the very least, the QCD-based assumptions must be scrapped and we have to resign ourselves to the incalculability of generic strong interactions~\cite{Lane:1994pg}. Said another way, the saturation of the correlation functions involved in S,T,U is unjustified -- one may expect a large, slowly varying coupling to result in a whole tower of (closely spaced) resonances, all contributing to low-energy observables. 

There has been some evidence that the near-conformal behavior of walking technicolor can also lead to a naturally reduced $S$-parameter~\cite{Kurachi:2006ej,Appelquist:1998xf, Appelquist:1999dq, Kurachi:2006mu}. The primary argument is that near-conformal behavior is accompanied by a near parity doubling of the resonance spectrum. Thus the techni-rho and techni-$a_T$ are nearly degenerate, as are their higher resonances. Simply from the definition of $S$ (Eq.~(\ref{eq:Slab})), one can see that a degenerate spectrum leads to small deviations in the SM electroweak sector~\footnote{For simple resonance models with zero or negative $S$ see Ref.~\cite{Knecht:1997ts}.}. 
The arguments for a degenerate resonance spectrum are based on numerical SDE results and by simple models of the effect of intermediate energy scales on $S$.

\subsection*{Walking and the top mass}

Based on SDE calculations, the maximum anomalous dimension a fermion bilinear can attain is $\gamma_m = 1$. However, even with maximal enhancement from walking, the large top quark mass is impossible to achieve with ETC scales consistent with Eq. (\ref{eq:FCNC1}, \ref{eq:FCNC2}). In order to accommodate the top mass, additional ingredients must be added to technicolor (this is why, in my opinion, it is the biggest problem with technicolor theories). Several mechanisms have been suggested to ameliorate this problem:
\begin{itemize}
\item Tumbling~\cite{Raby:1979my, Kikukawa:1992ig} technicolor theories: This idea suggests that there are different ETC scales for each generation: $\Lambda_{i}$ for generation $i$, with $\Lambda_1 \gg \Lambda_2 \gg \Lambda_3$. By separating the scales, we can keep $\Lambda_1$ consistent with Eq.(\ref{eq:FCNC1}, \ref{eq:FCNC2}), while lowering $\Lambda_3$ enough to generate realistic top masses. Unfortunately, it is difficult to build models which achieve multiple scales using gauge dynamics alone~\cite{Appelquist:2003hn,Appelquist:2004ai}. 
\item Topcolor-Assisted technicolor~\cite{Hill:1994hp}:  Another mechanism to enhance the top mass is to give the third generation separate, stronger $SU(3)_2$ and $U(1)_2$ gauge interactions. The $SU(3)_2$ (and $U(1)_2$) of the third generation and the $SU(3)_1, (U(1)_1)$ of the light generations are spontaneously broken at a scale $\sim 5\ \tev$ down to conventional color $SU(3)$ and hypercharge, leaving an octet of massive colored gauge bosons and a massive $Z'$. The heavy gauge bosons couple strongly to the third generation. If the coupling is strong enough, the dimension 6 four-fermion operator generated by heavy gauge boson exchange can cause a dynamical mass to form. The formalism which demonstrates the formation of a dynamical mass resums the four-fermion interactions to all orders, and is due to Nambu and Jona-Lasinio~\cite{Nambu:1961tp, Nambu:1961fr}. The exchange  of  the extra $U(1)$ gauge boson is arranged to be attractive for the top quark but repulsive for the bottom quark -- so we get an enhanced top mass without a large bottom mass. This usually requires either a strong $U(1)_2$ coupling, leading to a Landau pole below the ETC scale, or an extreme tuning in the coupling of the third generation's $SU(3)$ coupling.
\item Larger anomalous dimension: The limit $\gamma \simle 1$ is purely a result of the SDE approach. Forgetting this approximation, the actual limit on the anomalous dimension comes from unitarity and is $\gamma \simle 2$.  Depending on the other ETC parameters, anomalous dimensions of $\gamma \sim 1.5$ are sufficient to generate the top mass with a consistent ETC scale, so perhaps the SDE is just an insufficient tool. Models aimed at $\gamma > 1$ have been proposed, under the name `Strong ETC'~\cite{Chivukula:1990bc}, or, more recently, Conformal Technicolor~\cite{Luty:2004ye}.
\end{itemize}
\subsection*{Walking at low $N_C$ with different fermion representations}

 In the above efforts in walking technicolor, we have focused on keeping the fermions in the fundamental representation while increasing their number. This is not the only approach. In fact, a whole interesting class of models have recently been constructed which utilize a small $SU(N_{TC})$ technicolor gauge group, and a small number of technifermions in a large technicolor representation~\cite{Sannino:2004qp, Evans:2005pu, Dietrich:2005wk,Foadi:2007ue}. These theories have a minimal value of the perturbative $S$ parameter, which can be reduced either by walking dynamics (see above) or by contributions from a new `lepton' doublet. Depending on the $N_{TC}$ and $N_F$, the existence of such `leptons' is necessary to cancel the global $SU(2)_w$ Witten anomaly, however the hierarchy between the lepton and neutrino within the additional doublet necessary for a negative $S$ is left as a requirement of the ETC theory. For phenomenological studies see Ref.~\cite{Belyaev:2008yj}. Within this class of theories there has been considerable effort to map the conformal window -- the ranges of $N_F, N_C$ for various technifermion representations where the theory is believed to exhibit walking behavior~\cite{Dietrich:2006cm, Ryttov:2007sr}. There are also lattice efforts underway using different fermion representations~\cite{Catterall:2007yx, DeGrand:2008dh}.

\section{New Tools for Technicolor: Extra-dimensional Models}
\label{sec:xdim}
While walking technicolor offers a potential escape from FCNC and precision electroweak constraints, many results are simply incalculable. While incalculable models are difficult to rule out, they have fairly limited predictive power.

Recently much progress has been made describing strongly coupled gauge theories in $d$ dimensions by using weakly coupled `dual'  gravity theories in $d+1$ dimensions. This duality has been most rigorously proven between specific supersymmetric conformal $4D$ theory and a particular string theory sitting in an 5d AdS background. This duality, known as AdS/CFT~\cite{hep-th/9802109, hep-th/9802150}, has inspired a new effort in modelling strong dynamics based on using weakly coupled 5D models. The general thinking is that we don't necessarily need a complete duality to gain some new insight; perhaps the symmetry pattern and separation of scales present in AdS is sufficient to gain some insight into the behavior of important operators like $\langle \bar T T \rangle$.


The starting point for an AdS model of strong interactions is a slice of AdS space: an interval in the fifth coordinate, $z$ between two branes. The geometry in the interval is
\be
ds^2 = \frac{\ell^2_0}{z^2}(\eta_{\mu\nu}dx^{\mu}dx^{\nu} - dz^2),\ z \in (\ell_0, \ell_1)
\ee 
Without the presence of the branes, this geometry is invariant under scalings in $z$. Interpreting the fifth dimension as the energy/renormalization scale, the 5D model describes a theory which is scale invariant between two momentum scales, $\Lambda_{UV} = \frac{1}{\ell_0}$ and $\Lambda_{IR} = \frac{1}{\ell_1}$. The UV scale is the cutoff of the theory, while the IR scale is where conformal invariance is spontaneously broken.  The AdS geometry alone models the 4D near conformality, but we also need to include chiral symmetry. This is done by including gauge fields in the bulk of AdS; to model a 4D theory with $SU(2)_L\otimes SU(2)_R$ chiral symmetry we introduce $SU(2)_L\otimes SU(2)_R$ 5D gauge fields propagating in the interval~\footnote{We put $SU(2)_L \otimes SU(2)_R$ in the bulk rather than $SU(2)_L \otimes U(1)_Y$ because we want our low energy theory (after chiral symmetry breaking) to contain a custodial symmetry.}. The strength of the 5D interactions is set by the 5D gauge coupling $g_5$. To model chiral symmetry breaking we can include a bulk scalar field -- a 5D Higgs -- which interacts with the gauge fields and has a nonzero vev. By choosing the bulk Higgs couplings correctly, when the Higgs breaks chiral symmetry (down to $SU(2)_V$) it also breaks electroweak symmetry in the observed pattern. In the 4D interpretation, the bulk Higgs corresponds to the $\langle \bar T T \rangle$ condensate. Alternatively, it is possible to break chiral and electroweak symmetry in these models using boundary conditions alone. Models without a bulk Higgs are known as `Higgsless' models~\cite{Csaki:2003dt, Csaki:2003zu,Cacciapaglia:2004rb}.

The next step in the 5D model is to reduce it to a 4D theory. This is done by the usual technique of KK decomposition, i.e. by writing each 5D field as a sum over 4D fields times a profile in the fifth dimension. The profiles are solved for using the classical equations of motion accompanied by boundary conditions (which are necessary since we are working in a finite extra dimension). By restricting to classical EOM, we are making the assumption that the theory is weakly coupled and that quantum effects $\mathcal O(g_5^2)$ are negligible. A small 5D couplings corresponds to a 4D theory with a large number of colors (large-$N_C$). 

With the profiles determined, the fifth dimension can be integrated over and we are left with a series of 4D fields. Within each KK tower, all fields have the same quantum numbers but different 5D profiles and thus different mass. The lightest fields -- known as `zero-modes' -- are interpreted as the SM gauge fields $W^{\pm}, Z, \gamma$. The subsequent replicas of the SM fields are now the resonances of the strongly interacting theory. 

Using a 5D description we have modeled a strongly interaction theory, yet the observables we get at the end are different than what we get from 4D models. In a 4D model, we know the microscopic degrees of freedom -- the technifermion lagrangian -- but we know nothing about the resonance masses or interactions. In 5D the opposite is true; we have the masses and the couplings of the resonances, but the only clues we have about the fundamental theory are the chiral symmetry and large-$N_C$~\footnote{It is possible to recreate many of the aspects of a 5D model purely in a purely 4D language using deconstruction~\cite{Chivukula:2002ej, Foadi:2003xa, SekharChivukula:2004mu, Chivukula:2005bn,Chivukula:2006cg,He:2007ge} or Hidden Local Symmetry~\cite{Casalbuoni:1984hm, Casalbuoni:1986vq, Casalbuoni:1986vq, Casalbuoni:1995qt} techniques}. 

With the resonance masses and couplings determined, we can calculate observables like $S$. Unfortunately, in the simplest AdS-5D models we find $S$ is large and positive~\cite{Csaki:2003zu,Agashe:2003zs, Agashe:2007mc}. By examining the spectrum, we can get some idea as to why, after all this 5D work, we are back to square one. The spectrum -- the mass ordering of different parity excitations of the simplest AdS models is very similar to QCD, and the correlation functions involved in $S$ are indeed dominated by the lowest resonances. While it is certainly a useful tool, 5D modeling needs to be extended beyond simple AdS in order to provide useful insight into phenomenologically viable technicolor.

\section{Technicolor Phenomenology in the LHC Era}
\label{sec:LHCphen}

As we have seen, model building for strongly coupled models of EWSB is extremely difficult. The simplest approaches, either based on QCD-rescaling or 5D-AdS models do not yield viable models. In order to proceed, phenomenologists have adopted a new approach of effective lagrangians. Rather than attempt to write a complete high-energy theory, instead write down a consistent set of interesting interactions, imposing phenomenological constraints to limit the number and size of interactions. In the next section we explore two example phenomenological technicolor models and their LHC signals.

\subsection*{Low-Scale Technicolor}

To have consistent fermion masses without excessive FCNC, technicolor must be a walking theory, and therefore we need a lot of technimatter (matter drives the $\beta_0$ coefficient to smaller values). This technimatter can either be many fermions in the fundamental representation, or technifermions in multiple representations. Because each technidoublet contributes to the effective EWSB scale (Eq. (\ref{eq:TCEWSB})), there is a relation between the $F_T$ scale of each technifermion and $v = 246\ \gev$.  For $N_F$ technifermions in the same representation, the relation is  $v^2 = N F^2_T$. If the technifermions are in multiple represenstations the relation is generalized to $v^2 = \sum_i N_i F^2_{T,i}$, where we sum over all represenatations~\footnote{The compositeness scale dictated by (Eq. (\ref{eq:dimtrans})) depends on the representation so different representations have different $F_T$, with the largest technifermion representation having the largest $F_T$ scale. The simplest multi-representation model is Two-Scale technicolor~\cite{Lane:1989ej}, which postulates only two technicolor scales, $F_1$ and $F_2$.}. In either case, lots of technimatter implies a small $F_T$ scale. The class of walking theories which contains a low scale are known as Low-Scale Technicolor (LSTC) scenarios.

%

In a Low-Scale Technicolor scenario the resonances associated with the low scale will be light $\sim \mathcal O(200 - 500\ \gev)$. Assuming these resonances can be treated in isolation (that is, without considering the other resonances in the theory), the couplings of the low-scale resonances to fermions and gauge bosons are suppressed by the ratio of $F_i$ scales and the ratio of masses,
\be
\label{eq:gffV}
g_{ff\rho_T} \sim g_{ffW}\frac{F_1}{v}\frac{m_W}{M_{\rho_T}}.
\ee
This allows light resonances without violating direct detection bounds from LEP and the Tevatron. Additionally, lots of technimatter implies a large chiral symmetry in LSTC models and therefore a lot of techni-pions. These techni-pions can interact with the SM gauge bosons, fermions, and the TC resonances. Because of walking effects, the techni-pion mass is enhanced relative to the spin-1 techni-mesons ($\rho_T, a_T$). As a result, decay modes like $\rho_T \rightarrow \pi_T^+ \pi_T^-, a_T \rightarrow 3\pi_T$ are likely to be closed and the resonances are forced to decay into SM gauge bosons. The couplings to SM gauge bosons are small (see Eq.~(\ref{eq:gffV}), therefore the TC resonances in this scenario are strikingly narrow, $\Gamma_{\rho} \sim \mathcal O(.5\ \gev)$~\cite{Lane:1991qh}. The final assumption of LSTC is that the $\rho_T$ and $a_T$ are nearly degenerate~\cite{Eichten:2007sx}. This assumption is based on the idea, as explained in section~(\ref{sec:PEWwalk}) that the $S$ parameter is reduced in walking theories due to spectrum degeneracy.

The simplest and most explored LSTC scenario, known as the technicolor `straw-man' model postulates $\mathcal O(9)$ doublets, so $F_T \sim \frac{v}{3}$ .  Several Tevatron  studies of the `straw-man' model have been performed in the past, focusing on the modes $p\bar p \rightarrow \rho_T \rightarrow \pi_T W$~\cite{Eichten:1996dx, Eichten:1997yq, Lane:1999uh, Lane:2002sm, Lane:2006ak}. Due to the excessive $\bar t t$ background $\rho_T \rightarrow \pi_T W$ is impossible at the LHC. Instead, the best modes for discovery are $\rho_T \rightarrow W Z \rightarrow 3\ell + \nu, a_T \rightarrow \gamma W \rightarrow \ell \nu \gamma$ and $\omega_T \rightarrow \gamma Z \rightarrow \ell\ell \gamma$. These decay modes have multiple photons and leptons, thus milder backgrounds and better mass resolution. The discovery potential in these three modes at the LHC was investigated recently in Ref.~\cite{Brooijmans:2008se}. The signals are striking -- all of these resonances could be discovered within the first few $\fb^{-1}$ of LHC running.
\begin{figure}[th!]
\centering
\begin{minipage}[c]{0.30\linewidth}
   \centering
   \includegraphics[width = 1.9 in, height=1.9 in]{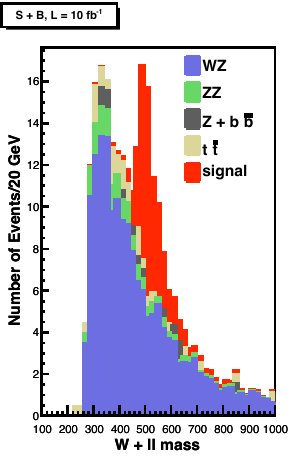} 
\end{minipage}
\hspace{0.5 cm}
\begin{minipage}[c]{0.3\linewidth}
   \centering
   \includegraphics[width = 1.9 in, height=1.9in]{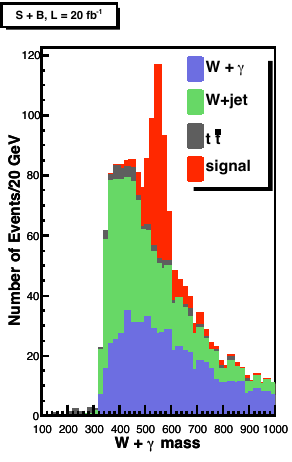}
\end{minipage}  
\hspace{0.5cm}
\begin{minipage}[c]{0.3\linewidth}
   \centering
   \includegraphics[width = 1.9 in, height=1.9in]{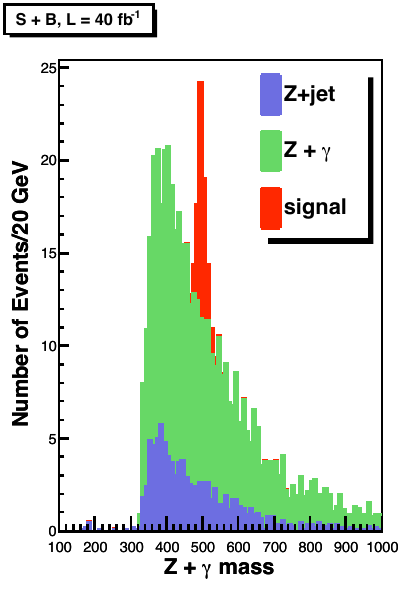}
  \end{minipage}  
\flushleft
\caption{LSTC signal plus SM background in the $WZ, W\gamma, {\rm and\ } Z\gamma$} channels. The signal in each channel is dominated by a single resonance: $\rho_T$ in $WZ$, $a_T$ in $W\gamma$ and $\omega_T$ in $Z\gamma$. These resonances could be discovered within the first few $\fb^{-1}$ of LHC running. The neutrino momentum in the $WZ$ and $W\gamma$ channels is reconstructed by the reqiurement that $M_{(p_{\ell} + p_{\nu})} = M_W$. The details of the analysis leading to these plots can be found in Ref.~\cite{Brooijmans:2008se}.
\end{figure}

Resonance searches in the modes $3\ell + \nu, \ell\nu\gamma, {\rm and\ } \ell\ell\gamma$ have received little attention, partially because they have no SUSY analogue. In the MSSM one could imagine searches for $H^{\pm}$ in the $3\ell + \nu, {\rm or\ }\ell\nu\gamma$ channels, however, the $H^{\pm}W^{\pm}Z$, $H^{\pm}\gamma W^{\pm}$ interactions vanish at tree level in the MSSM (or in any other two-Higgs doublet model).

\subsection*{Holographic Technicolor}

Until recently, all modern technicolor collider studies were done in the LSTC context~\footnote{ The reason so many studies exist for the `straw-man' technicolor model is that it is currently the only technicolor model implemented in the ubiquitous Monte Carlo program PYTHIA.}. While these are important studies, it is also important to explore other effective theories and frameworks of EW-scale strong interactions. 

To  study a wider array of technicolor models and access phenomenology not present in the LSTC, a variant of the Higgsless setup, known as Holographic Technicolor (HTC) was developed in Ref.~\cite{Hirn:2006nt, Hirn:2006wg} . The setup of HTC is identical to the Higgsless setup reviewed in section (\ref{sec:xdim}), but with one important difference -- the assumption that all bulk fields feel the same 5D geometry has been dropped. Instead, HTC assumes a more general action
\begin{gather}
\mathcal L = -\frac{1}{g^2_5}\int d^5x~ \omega_V(z) (F_{V, NM}F^{NM}_V) + \omega_A(z)(F_{A, NM}F^{NM}_A) \\
 \omega_{V,A}(z) = \frac{\ell_0}{z}\exp{o_{V,A}\Big( \frac{z}{\ell_1}\Big)^4}, \quad o_{V,A} \simle 0 \nonumber
\end{gather}
The extra parameters in HTC relative to traditional Higgsless are the geometry parameters $o_{V,A}$. Nonzero values for these parameters indicate a deviation from AdS in the IR which grows stronger with increasing $z$. The exact power of $4$ is unimportant; a different power of $z$ can be compensated with a different $o_{V,A}$ and will lead to the same physics. In practice $\omega_V \ne \omega_A$ allows us to dial the properties of the axial resonances independently of the vector resonances. This is not possible in LSTC, where a techni-parity symmetry is assumed. Actually, assuming a techni-parity symmetry is  artificial since the electroweak interactions {\em automatically} break parity. By divorcing vector from axial, we can study a much wider subset of technicolor scenarios. A few benchmark scenarios which focused on $M_{\rho_T} \cong M_{a_T}$ for precision electroweak considerations~\cite{Hirn:2006nt}, were studied in~\cite{Hirn:2007we,Hirn:2008tc}.  The key difference between Ref.~\cite{Hirn:2007we,Hirn:2008tc} and LSTC is that both the $\rho_T$ and $a_T$ can couple to $WZ, W\gamma$, resulting in the distinct two-peak feature seen below. In LSTC, the $a_TWZ$ and $\rho_T WZ$ couplings are suppressed by techniparity.
\begin{figure}[th!]
\centering
\begin{minipage}[c]{0.30\linewidth}
   \centering
   \includegraphics[width = 1.9 in, height=1.9 in]{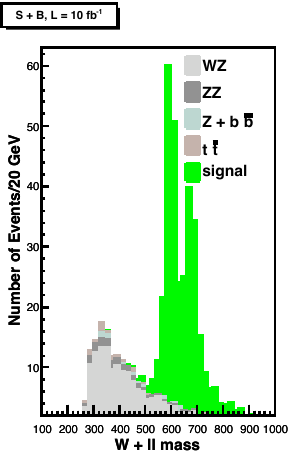} 
 \end{minipage}
\hspace{0.5 cm}
\begin{minipage}[c]{0.3\linewidth}
   \centering
   \includegraphics[width = 1.9 in, height=1.9in]{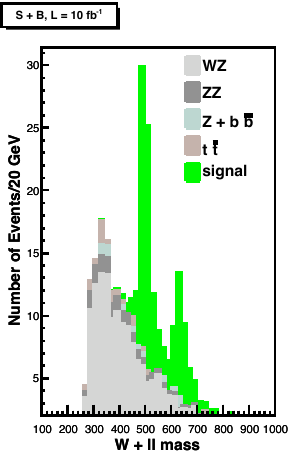}
 \end{minipage}  
\hspace{0.5cm}
\begin{minipage}[c]{0.3\linewidth}
   \centering
   \includegraphics[width = 1.9 in, height=1.9in]{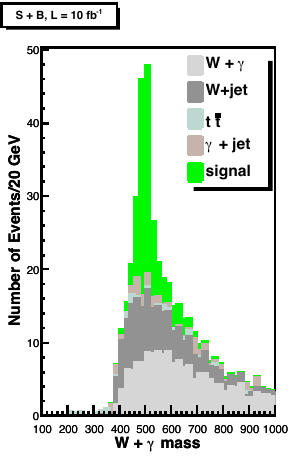}
  \end{minipage}
  \caption{Possible resonance signals in the $WZ$ channel (left two figures), or $W\gamma$ channel (rightmost figure) generated within the HTC framework. By dialing the geometry parameters $o_V, o_A$ we recreate a scenario where {\em both} the two lightest charged resonance KK states couple to $WZ$ or $W\gamma$, leading to the striking two-peak feature. For the details of the parameter choices in these plots see Ref.~\cite{Hirn:2007we} }  
\end{figure}
More complicated processes, such as vector boson fusion $pp \rightarrow W/Z + jj$, which probes the resonance effects on unitarizing gauge boson scattering, and associated production $pp\rightarrow W^* \rightarrow W/Z + \rho_T$ have also been studied. These processes, which involve more final state particles and elaborate cuts to remove SM backgrounds, often require high luminosity $(100 - 300\ \fb^{-1})$ for discovery.

\section{Conclusions}
The goal of the LHC is to unravel the mechanism behind electroweak symmetry breaking. We should be open to all options and not just those with pretty, but irrelevant aspects. Technicolor is the {\em only} completely natural solution to the gauge hierarchy problem, however it involves strong coupling. I view this as an interesting, compelling challenge rather than a deterrent. There is, as yet, no evidence for Technicolor. However,  as with supersymmetry or most other schemes for beyond the SM physics -- technicolor cannot be ruled out. To ameliorate the severe flavor constraints, modern technicolor is believed to have a walking (nearly conformal) coupling over a wide range of energies. This departure from familiar strong interactions makes model building challenging, and phenomenologists have turned to effective theories for collider studies. While not fundamental, phenomenological models of strong interactions describe a host of striking features which are well within the grasp of the LHC. 

\section*{Acknowledgements}
I would like to thank the organizers Professors Antonino Zichichi and Gerard 't Hooft for the invitation to attend and lecture at this school. I would also like to thank Prof. Kenneth Lane for recommending me to the organizers and for his guidance in preparing the lecture and notes.  AM is supported by DOE grant DE-FG02-92ER-40704.

\bibliography{aomartin.bib}
\bibliographystyle{utcaps}

\end{document}

%% file: defs.tex
\newcommand\ltap{\
  \raise.3ex\hbox{$<$\kern-.75em\lower1ex\hbox{$\sim$}}\ } 
\newcommand\gtap{\
  \raise.3ex\hbox{$>$\kern-.75em\lower1ex\hbox{$\sim$}}\ } 

\newcommand\simge{\mathrel{%
   \rlap{\raise 0.511ex \hbox{$>$}}{\lower 0.511ex \hbox{$\sim$}}}}
\newcommand\simle{\mathrel{
   \rlap{\raise 0.511ex \hbox{$<$}}{\lower 0.511ex \hbox{$\sim$}}}}

\newcommand{\slashchar}[1]%
        {\kern .25em\raise.18ex\hbox{$/$}\kern-.75em #1}
\def\lsim{\mathrel{\raise.3ex\hbox{$<$\kern-.75em\lower1ex\hbox{$\sim$}}}}
\def\gsim{\mathrel{\raise.3ex\hbox{$>$\kern-.75em\lower1ex\hbox{$\sim$}}}}

\newcommand\be{\begin{equation}} 
\newcommand\ee{\end{equation}} 
\newcommand\bea{\begin{eqnarray}}
\newcommand\eea{\end{eqnarray}}
\newcommand\ba{\begin{array}}
\newcommand\ea{\end{array}}
\newcommand\nn{\nonumber}

\newcommand\mev{{\rm MeV}}
\newcommand\gev{{\rm GeV}}
\newcommand\tev{{\rm TeV}}

\newcommand\fb{{\rm fb}}

%% file: aomartin_TC_Erice_notes.bbl
\providecommand{\href}[2]{#2}\begingroup\raggedright\begin{thebibliography}{10}

\bibitem{Lane:1993wz}
K.~D. Lane, ``{An Introduction to technicolor},''
  \href{http://xxx.lanl.gov/abs/hep-ph/9401324}{ hep-ph/9401324}.

\bibitem{Lane:1994vu}
K.~D. Lane, ``{Technicolor},'' \href{http://xxx.lanl.gov/abs/hep-ph/9501249}{
  hep-ph/9501249}.

\bibitem{Chivukula:1998if}
R.~S. Chivukula, ``{Models of electroweak symmetry breaking},''
  \href{http://xxx.lanl.gov/abs/hep-ph/9803219}{ hep-ph/9803219}.

\bibitem{Lane:2000pa}
K.~D. Lane, ``{Technicolor 2000},''
  \href{http://xxx.lanl.gov/abs/hep-ph/0007304}{ hep-ph/0007304}.

\bibitem{Lane:2002wv}
K.~Lane, ``{Two lectures on technicolor},''
  \href{http://xxx.lanl.gov/abs/hep-ph/0202255}{ hep-ph/0202255}.

\bibitem{Hill:2002ap}
C.~T. Hill and E.~H. Simmons, ``{Strong dynamics and electroweak symmetry
  breaking},'' {\em Phys. Rept.} {\bf 381} (2003) 235--402,
  \href{http://xxx.lanl.gov/abs/hep-ph/0203079}{ hep-ph/0203079}.

\bibitem{Wilson:1971dh}
K.~G. Wilson, ``{Renormalization group and critical phenomena. 2. Phase space
  cell analysis of critical behavior},'' {\em Phys. Rev.} {\bf B4} (1971)
  3184--3205.

\bibitem{Wilson:1973jj}
K.~G. Wilson and J.~B. Kogut, ``{The Renormalization group and the epsilon
  expansion},'' {\em Phys. Rept.} {\bf 12} (1974) 75--200.

\bibitem{Bardeen:1957mv}
J.~Bardeen, L.~N. Cooper, and J.~R. Schrieffer, ``{Theory of
  superconductivity},'' {\em Phys. Rev.} {\bf 108} (1957) 1175--1204.

\bibitem{Weinberg:1979bn}
S.~Weinberg, ``Implications of Dynamical Symmetry Breaking: an addendum,'' {\em
  Phys. Rev.} {\bf D19} (1979) 1277--1280.

\bibitem{Susskind:1978ms}
L.~Susskind, ``Dynamics of Spontaneous Symmetry Breaking in the Weinberg-Salam
  Theory,'' {\em Phys. Rev.} {\bf D20} (1979) 2619--2625.

\bibitem{Eichten:1984eu}
E.~Eichten, I.~Hinchliffe, K.~D. Lane, and C.~Quigg, ``Super Collider
  Physics,'' {\em Rev. Mod. Phys.} {\bf 56} (1984) 579--707.

\bibitem{Bagger:1993zf}
J.~Bagger {\em et.~al.}, ``{The Strongly interacting W W system: Gold plated
  modes},'' {\em Phys. Rev.} {\bf D49} (1994) 1246--1264,
  \href{http://xxx.lanl.gov/abs/hep-ph/9306256}{ hep-ph/9306256}.

\bibitem{Bagger:1995mk}
J.~Bagger {\em et.~al.}, ``{CERN LHC analysis of the strongly interacting W W
  system: Gold plated modes},'' {\em Phys. Rev.} {\bf D52} (1995) 3878--3889,
  \href{http://xxx.lanl.gov/abs/hep-ph/9504426}{ hep-ph/9504426}.

\bibitem{Eichten:1979ah}
E.~Eichten and K.~D. Lane, ``{Dynamical Breaking of Weak Interaction
  Symmetries},'' {\em Phys. Lett.} {\bf B90} (1980) 125--130.

\bibitem{Peskin:1990zt}
M.~E. Peskin and T.~Takeuchi, ``A new constraint on a strongly interacting
  Higgs sector,'' {\em Phys. Rev. Lett.} {\bf 65} (1990) 964--967.

\bibitem{Golden:1990ig}
M.~Golden and L.~Randall, ``Radiative corrections to electroweak parameters in
  technicolor theories,'' {\em Nucl. Phys.} {\bf B361} (1991) 3--23.

\bibitem{Amsler:2008zz}
{\bf Particle Data Group} Collaboration, C.~Amsler {\em et.~al.}, ``{Review of
  particle physics},'' {\em Phys. Lett.} {\bf B667} (2008) 1.

\bibitem{Appelquist:2007hu}
T.~Appelquist, G.~T. Fleming, and E.~T. Neil, ``{Lattice Study of the Conformal
  Window in QCD-like Theories},'' {\em Phys. Rev. Lett.} {\bf 100} (2008)
  171607, \href{http://xxx.lanl.gov/abs/0712.0609}{ 0712.0609}.

\bibitem{Catterall:2007yx}
S.~Catterall and F.~Sannino, ``{Minimal walking on the lattice},'' {\em Phys.
  Rev.} {\bf D76} (2007) 034504, \href{http://xxx.lanl.gov/abs/0705.1664}{
  0705.1664}.

\bibitem{Deuzeman:2008da}
A.~Deuzeman, M.~P. Lombardo, and E.~Pallante, ``{Hunting for the Conformal
  Window},'' \href{http://xxx.lanl.gov/abs/0810.3117}{ 0810.3117}.

\bibitem{Fodor:2008hn}
Z.~Fodor, K.~Holland, J.~Kuti, D.~Nogradi, and C.~Schroeder, ``{Probing
  technicolor theories with staggered fermions},''
  \href{http://xxx.lanl.gov/abs/0809.4890}{ 0809.4890}.

\bibitem{Fodor:2008hm}
Z.~Fodor, K.~Holland, J.~Kuti, D.~Nogradi, and C.~Schroeder, ``{Nearly
  conformal electroweak sector with chiral fermions},''
  \href{http://xxx.lanl.gov/abs/0809.4888}{ 0809.4888}.

\bibitem{Holdom:1981rm}
B.~Holdom, ``Raising the sideways scale,'' {\em Phys. Rev.} {\bf D24} (1981)
  1441.

\bibitem{Appelquist:1986an}
T.~W. Appelquist, D.~Karabali, and L.~C.~R. Wijewardhana, ``Chiral Hierarchies
  and the Flavor Changing Neutral Current Problem in Technicolor,'' {\em Phys.
  Rev. Lett.} {\bf 57} (1986) 957.

\bibitem{Yamawaki:1986zg}
K.~Yamawaki, M.~Bando, and K.-i. Matumoto, ``Scale invariant technicolor model
  and a technidilaton,'' {\em Phys. Rev. Lett.} {\bf 56} (1986) 1335.

\bibitem{Pagels:1974se}
H.~Pagels, ``{Departures from Chiral Symmetry: A Review},'' {\em Phys. Rept.}
  {\bf 16} (1975) 219.

\bibitem{Fukuda:1976zb}
R.~Fukuda and T.~Kugo, ``{Schwinger-Dyson Equation for Massless Vector Theory
  and Absence of Fermion Pole},'' {\em Nucl. Phys.} {\bf B117} (1976) 250.

\bibitem{Higashijima:1983gx}
K.~Higashijima, ``{Dynamical Chiral Symmetry Breaking},'' {\em Phys. Rev.} {\bf
  D29} (1984) 1228.

\bibitem{Appelquist:1988yc}
T.~Appelquist, K.~D. Lane, and U.~Mahanta, ``{ON THE LADDER APPROXIMATION FOR
  SPONTANEOUS CHIRAL SYMMETRY BREAKING},'' {\em Phys. Rev. Lett.} {\bf 61}
  (1988) 1553.

\bibitem{Cohen:1988sq}
A.~G. Cohen and H.~Georgi, ``{WALKING BEYOND THE RAINBOW},'' {\em Nucl. Phys.}
  {\bf B314} (1989) 7.

\bibitem{Kurachi:2006ej}
M.~Kurachi and R.~Shrock, ``{Study of the change from walking to non-walking
  behavior in a vectorial gauge theory as a function of N(f)},'' {\em JHEP}
  {\bf 12} (2006) 034, \href{http://xxx.lanl.gov/abs/hep-ph/0605290}{
  hep-ph/0605290}.

\bibitem{Lane:1994pg}
K.~D. Lane, ``{Technicolor and precision tests of the electroweak
  interactions},'' \href{http://xxx.lanl.gov/abs/hep-ph/9409304}{
  hep-ph/9409304}.

\bibitem{Appelquist:1998xf}
T.~Appelquist and F.~Sannino, ``The physical spectrum of conformal SU(N) gauge
  theories,'' {\em Phys. Rev.} {\bf D59} (1999) 067702,
  \href{http://xxx.lanl.gov/abs/hep-ph/9806409}{ hep-ph/9806409}.

\bibitem{Appelquist:1999dq}
T.~Appelquist, P.~S. Rodrigues~da Silva, and F.~Sannino, ``{Enhanced global
  symmetries and the chiral phase transition},'' {\em Phys. Rev.} {\bf D60}
  (1999) 116007, \href{http://xxx.lanl.gov/abs/hep-ph/9906555}{
  hep-ph/9906555}.

\bibitem{Kurachi:2006mu}
M.~Kurachi and R.~Shrock, ``{Behavior of the S parameter in the crossover
  region between walking and QCD-like regimes of an SU(N) gauge theory},'' {\em
  Phys. Rev.} {\bf D74} (2006) 056003,
  \href{http://xxx.lanl.gov/abs/hep-ph/0607231}{ hep-ph/0607231}.

\bibitem{Raby:1979my}
S.~Raby, S.~Dimopoulos, and L.~Susskind, ``{Tumbling Gauge Theories},'' {\em
  Nucl. Phys.} {\bf B169} (1980) 373.

\bibitem{Kikukawa:1992ig}
Y.~Kikukawa and N.~Kitazawa, ``{Tumbling and technicolor theory},'' {\em Phys.
  Rev.} {\bf D46} (1992) 3117--3122.

\bibitem{Appelquist:2003hn}
T.~Appelquist, M.~Piai, and R.~Shrock, ``{Fermion masses and mixing in extended
  technicolor models},'' {\em Phys. Rev.} {\bf D69} (2004) 015002,
  \href{http://xxx.lanl.gov/abs/hep-ph/0308061}{ hep-ph/0308061}.

\bibitem{Appelquist:2004ai}
T.~Appelquist, N.~D. Christensen, M.~Piai, and R.~Shrock, ``{Flavor-changing
  processes in extended technicolor},'' {\em Phys. Rev.} {\bf D70} (2004)
  093010, \href{http://xxx.lanl.gov/abs/hep-ph/0409035}{ hep-ph/0409035}.

\bibitem{Hill:1994hp}
C.~T. Hill, ``Topcolor assisted technicolor,'' {\em Phys. Lett.} {\bf B345}
  (1995) 483--489, \href{http://xxx.lanl.gov/abs/hep-ph/9411426}{
  hep-ph/9411426}.

\bibitem{Nambu:1961tp}
Y.~Nambu and G.~Jona-Lasinio, ``{Dynamical model of elementary particles based
  on an analogy with superconductivity. I},'' {\em Phys. Rev.} {\bf 122} (1961)
  345--358.

\bibitem{Nambu:1961fr}
Y.~Nambu and G.~Jona-Lasinio, ``{Dynamical model of elementary particles based
  on an analogy with superconductivity. II},'' {\em Phys. Rev.} {\bf 124}
  (1961) 246--254.

\bibitem{Chivukula:1990bc}
R.~S. Chivukula, A.~G. Cohen, and K.~D. Lane, ``{ASPECTS OF DYNAMICAL
  ELECTROWEAK SYMMETRY BREAKING},'' {\em Nucl. Phys.} {\bf B343} (1990)
  554--570.

\bibitem{Luty:2004ye}
M.~A. Luty and T.~Okui, ``{Conformal technicolor},'' {\em JHEP} {\bf 09} (2006)
  070, \href{http://xxx.lanl.gov/abs/hep-ph/0409274}{ hep-ph/0409274}.

\bibitem{Sannino:2004qp}
F.~Sannino and K.~Tuominen, ``{Techniorientifold},'' {\em Phys. Rev.} {\bf D71}
  (2005) 051901, \href{http://xxx.lanl.gov/abs/hep-ph/0405209}{
  hep-ph/0405209}.

\bibitem{Evans:2005pu}
N.~Evans and F.~Sannino, ``{Minimal walking technicolour, the top mass and
  precision electroweak measurements},''
  \href{http://xxx.lanl.gov/abs/hep-ph/0512080}{ hep-ph/0512080}.

\bibitem{Dietrich:2005wk}
D.~D. Dietrich, F.~Sannino, and K.~Tuominen, ``{Light composite Higgs and
  precision electroweak measurements on the Z resonance: An update},'' {\em
  Phys. Rev.} {\bf D73} (2006) 037701,
  \href{http://xxx.lanl.gov/abs/hep-ph/0510217}{ hep-ph/0510217}.

\bibitem{Foadi:2007ue}
R.~Foadi, M.~T. Frandsen, T.~A. Ryttov, and F.~Sannino, ``{Minimal Walking
  Technicolor: Set Up for Collider Physics},'' {\em Phys. Rev.} {\bf D76}
  (2007) 055005, \href{http://xxx.lanl.gov/abs/0706.1696}{ 0706.1696}.

\bibitem{Belyaev:2008yj}
A.~Belyaev {\em et.~al.}, ``{Technicolor Walks at the LHC},''
  \href{http://xxx.lanl.gov/abs/0809.0793}{ 0809.0793}.

\bibitem{Dietrich:2006cm}
D.~D. Dietrich and F.~Sannino, ``{Walking in the SU(N)},'' {\em Phys. Rev.}
  {\bf D75} (2007) 085018, \href{http://xxx.lanl.gov/abs/hep-ph/0611341}{
  hep-ph/0611341}.

\bibitem{Ryttov:2007sr}
T.~A. Ryttov and F.~Sannino, ``{Conformal Windows of SU(N) Gauge Theories,
  Higher Dimensional Representations and The Size of The Unparticle World},''
  {\em Phys. Rev.} {\bf D76} (2007) 105004,
  \href{http://xxx.lanl.gov/abs/0707.3166}{ 0707.3166}.

\bibitem{DeGrand:2008dh}
T.~DeGrand, Y.~Shamir, and B.~Svetitsky, ``{Exploring the phase diagram of
  sextet QCD},'' \href{http://xxx.lanl.gov/abs/0809.2953}{ 0809.2953}.

\bibitem{hep-th/9802109}
S.~S. Gubser, I.~R. Klebanov, and A.~M. Polyakov, ``Gauge theory correlators
  from non-critical string theory,'' {\em Phys. Lett.} {\bf B428} (1998)
  105--114, \href{http://xxx.lanl.gov/abs/hep-th/9802109}{ hep-th/9802109}.

\bibitem{hep-th/9802150}
E.~Witten, ``Anti-de Sitter space and holography,'' {\em Adv. Theor. Math.
  Phys.} {\bf 2} (1998) 253--291,
  \href{http://xxx.lanl.gov/abs/hep-th/9802150}{ hep-th/9802150}.

\bibitem{Csaki:2003dt}
C.~Csaki, C.~Grojean, H.~Murayama, L.~Pilo, and J.~Terning, ``Gauge theories on
  an interval: Unitarity without a Higgs,'' {\em Phys. Rev.} {\bf D69} (2004)
  055006, \href{http://xxx.lanl.gov/abs/hep-ph/0305237}{ hep-ph/0305237}.

\bibitem{Csaki:2003zu}
C.~Csaki, C.~Grojean, L.~Pilo, and J.~Terning, ``Towards a realistic model of
  Higgsless electroweak symmetry breaking,'' {\em Phys. Rev. Lett.} {\bf 92}
  (2004) 101802, \href{http://xxx.lanl.gov/abs/hep-ph/0308038}{
  hep-ph/0308038}.

\bibitem{Cacciapaglia:2004rb}
G.~Cacciapaglia, C.~Csaki, C.~Grojean, and J.~Terning, ``Curing the ills of
  Higgsless models: The S parameter and unitarity,'' {\em Phys. Rev.} {\bf D71}
  (2005) 035015, \href{http://xxx.lanl.gov/abs/hep-ph/0409126}{
  hep-ph/0409126}.

\bibitem{Agashe:2003zs}
K.~Agashe, A.~Delgado, M.~J. May, and R.~Sundrum, ``RS1, custodial isospin and
  precision tests,'' {\em JHEP} {\bf 08} (2003) 050,
  \href{http://xxx.lanl.gov/abs/hep-ph/0308036}{ hep-ph/0308036}.

\bibitem{Agashe:2007mc}
K.~Agashe, C.~Csaki, C.~Grojean, and M.~Reece, ``{The S-parameter in
  holographic technicolor models},'' {\em JHEP} {\bf 12} (2007) 003,
  \href{http://xxx.lanl.gov/abs/0704.1821}{ 0704.1821}.

\bibitem{Lane:1991qh}
K.~D. Lane and M.~V. Ramana, ``{Walking technicolor signatures at hadron
  colliders},'' {\em Phys. Rev.} {\bf D44} (1991) 2678--2700.

\bibitem{Eichten:2007sx}
E.~Eichten and K.~Lane, ``{Low-scale technicolor at the Tevatron and LHC},''
  \href{http://xxx.lanl.gov/abs/0706.2339}{ 0706.2339}.

\bibitem{Eichten:1996dx}
E.~Eichten and K.~D. Lane, ``{Low-scale technicolor at the Tevatron},'' {\em
  Phys. Lett.} {\bf B388} (1996) 803--807,
  \href{http://xxx.lanl.gov/abs/hep-ph/9607213}{ hep-ph/9607213}.

\bibitem{Eichten:1997yq}
E.~Eichten, K.~D. Lane, and J.~Womersley, ``{Finding low-scale technicolor at
  hadron colliders},'' {\em Phys. Lett.} {\bf B405} (1997) 305--311,
  \href{http://xxx.lanl.gov/abs/hep-ph/9704455}{ hep-ph/9704455}.

\bibitem{Lane:1999uh}
K.~D. Lane, ``{Technihadron production and decay in low-scale technicolor},''
  {\em Phys. Rev.} {\bf D60} (1999) 075007,
  \href{http://xxx.lanl.gov/abs/hep-ph/9903369}{ hep-ph/9903369}.

\bibitem{Lane:2002sm}
K.~Lane and S.~Mrenna, ``{The collider phenomenology of technihadrons in the
  technicolor Straw Man Model},'' {\em Phys. Rev.} {\bf D67} (2003) 115011,
  \href{http://xxx.lanl.gov/abs/hep-ph/0210299}{ hep-ph/0210299}.

\bibitem{Lane:2006ak}
K.~Lane, ``{Search for low-scale technicolor at the Tevatron},''
  \href{http://xxx.lanl.gov/abs/hep-ph/0605119}{ hep-ph/0605119}.

\bibitem{Brooijmans:2008se}
G.~Brooijmans {\em et.~al.}, ``{New Physics at the LHC: A Les Houches Report.
  Physics at Tev Colliders 2007 -- New Physics Working Group},''
  \href{http://xxx.lanl.gov/abs/0802.3715}{ 0802.3715}.

\bibitem{Hirn:2006nt}
J.~Hirn and V.~Sanz, ``A negative S parameter from holographic technicolor,''
  {\em Phys. Rev. Lett.} {\bf 97} (2006) 121803,
  \href{http://xxx.lanl.gov/abs/hep-ph/0606086}{ hep-ph/0606086}.

\bibitem{Hirn:2006wg}
J.~Hirn and V.~Sanz, ``The fifth dimension as an analogue computer for strong
  interactions at the LHC,'' \href{http://xxx.lanl.gov/abs/hep-ph/0612239}{
  hep-ph/0612239}.

\bibitem{Hirn:2007we}
J.~Hirn, A.~Martin, and V.~Sanz, ``{Benchmarks for new strong interactions at
  the LHC},'' {\em JHEP} {\bf 05} (2008) 084,
  \href{http://xxx.lanl.gov/abs/0712.3783}{ 0712.3783}.

\bibitem{Hirn:2008tc}
J.~Hirn, A.~Martin, and V.~Sanz, ``{Describing viable technicolor scenarios},''
  {\em Phys. Rev.} {\bf D78} (2008) 075026,
  \href{http://xxx.lanl.gov/abs/0807.2465}{ 0807.2465}.

\bibitem{Knecht:1997ts}
M.~Knecht and E.~de~Rafael, ``Patterns of spontaneous chiral symmetry breaking
  in the large N(c) limit of QCD-like theories,'' {\em Phys. Lett.} {\bf B424}
  (1998) 335--342, \href{http://xxx.lanl.gov/abs/hep-ph/9712457}{
  hep-ph/9712457}.

\bibitem{Chivukula:2002ej}
R.~S. Chivukula and H.-J. He, ``{Unitarity of deconstructed five-dimensional
  Yang-Mills theory},'' {\em Phys. Lett.} {\bf B532} (2002) 121--128,
  \href{http://xxx.lanl.gov/abs/hep-ph/0201164}{ hep-ph/0201164}.

\bibitem{Foadi:2003xa}
R.~Foadi, S.~Gopalakrishna, and C.~Schmidt, ``{Higgsless electroweak symmetry
  breaking from theory space},'' {\em JHEP} {\bf 03} (2004) 042,
  \href{http://xxx.lanl.gov/abs/hep-ph/0312324}{ hep-ph/0312324}.

\bibitem{SekharChivukula:2004mu}
R.~Sekhar~Chivukula, E.~H. Simmons, H.-J. He, M.~Kurachi, and M.~Tanabashi,
  ``Electroweak corrections and unitarity in linear moose models,'' {\em Phys.
  Rev.} {\bf D71} (2005) 035007, \href{http://xxx.lanl.gov/abs/hep-ph/0410154}{
  hep-ph/0410154}.

\bibitem{Chivukula:2005bn}
R.~S. Chivukula, E.~H. Simmons, H.-J. He, M.~Kurachi, and M.~Tanabashi,
  ``Deconstructed Higgsless models with one-site delocalization,'' {\em Phys.
  Rev.} {\bf D71} (2005) 115001, \href{http://xxx.lanl.gov/abs/hep-ph/0502162}{
  hep-ph/0502162}.

\bibitem{Chivukula:2006cg}
R.~S. Chivukula {\em et.~al.}, ``{A three site higgsless model},'' {\em Phys.
  Rev.} {\bf D74} (2006) 075011, \href{http://xxx.lanl.gov/abs/hep-ph/0607124}{
  hep-ph/0607124}.

\bibitem{He:2007ge}
H.-J. He {\em et.~al.}, ``{LHC Signatures of New Gauge Bosons in Minimal
  Higgsless Model},'' {\em Phys. Rev.} {\bf D78} (2008) 031701,
  \href{http://xxx.lanl.gov/abs/0708.2588}{ 0708.2588}.

\bibitem{Casalbuoni:1984hm}
R.~Casalbuoni, D.~Dominici, and R.~Gatto, ``{EFFECTIVE LAGRANGIAN DESCRIPTION
  OF THE POSSIBLE STRONG SECTOR OF THE STANDARD MODEL},'' {\em Phys. Lett.}
  {\bf B147} (1984) 419.

\bibitem{Casalbuoni:1986vq}
R.~Casalbuoni, S.~De~Curtis, D.~Dominici, and R.~Gatto, ``{Physical
  Implications of Possible J=1 Bound States from Strong Higgs},'' {\em Nucl.
  Phys.} {\bf B282} (1987) 235.

\bibitem{Casalbuoni:1995qt}
R.~Casalbuoni {\em et.~al.}, ``{Degenerate BESS Model: The possibility of a low
  energy strong electroweak sector},'' {\em Phys. Rev.} {\bf D53} (1996)
  5201--5221, \href{http://xxx.lanl.gov/abs/hep-ph/9510431}{ hep-ph/9510431}.

\bibitem{Lane:1989ej}
K.~D. Lane and E.~Eichten, ``{Two Scale Technicolor},'' {\em Phys. Lett.} {\bf
  B222} (1989) 274.

\end{thebibliography}\endgroup
